\documentclass[aps,twocolumn,showpacs,superscriptaddress,floatfix,prb]{revtex4-2}

\usepackage{amsmath}
\usepackage{amssymb}
\usepackage{graphicx}
\usepackage{hyperref}
\usepackage{color}
\usepackage{bm}
\usepackage{braket}
\usepackage{siunitx}
\usepackage{booktabs}
\usepackage{adjustbox}
\usepackage[capitalise]{cleveref} 
\hypersetup{hypertex=true,colorlinks=true,linkcolor=blue,anchorcolor=blue,citecolor=blue}

\newcommand{\C}{\mathrm{C}}

\begin{document}

\title{Efficient iPEPS Simulation on the Honeycomb Lattice via QR-based CTMRG}

\author{Qi Yang}
\affiliation{Institute for Theoretical Physics, University of Amsterdam, Science Park 904, 1098 XH Amsterdam, The Netherlands}
\author{Philippe Corboz}
\affiliation{Institute for Theoretical Physics, University of Amsterdam, Science Park 904, 1098 XH Amsterdam, The Netherlands}

\date{\today}

\begin{abstract}
	We develop a QR-based corner transfer matrix renormalization group (CTMRG) framework for contracting infinite projected entangled-pair states (iPEPS) on honeycomb lattices. Our method explicitly uses the lattice's native $\C_{3v}$ symmetry at each site, generalizing QR-based acceleration (previously limited to square lattices) to enable efficient and stable contractions. This approach achieves order-of-magnitude speedups over conventional singular value decomposition (SVD)-based CTMRG while maintaining high numerical precision. Comprehensive benchmark calculations for the spin-1/2 Heisenberg and Kitaev models demonstrate higher computational efficiency without sacrificing accuracy. We further employ our method to study the Kitaev-Heisenberg model, where we provide numerical evidence for the universal $1/r^4$ decay of the dimer-dimer correlation function within the quantum spin liquid (QSL) phase. Our work establishes a framework for extending QR-based CTMRG to other lattice geometries, opening new avenues for studying exotic quantum phases with tensor networks.
\end{abstract}

\maketitle

\section{Introduction}
The honeycomb lattice has proven to be an exceptionally rich platform for investigating exotic quantum phases in two-dimensional (2D) systems, including magnetic states, valence bond solids, and quantum spin liquid behavior~\cite{kitaev2006anyons,PhysRevLett.105.027204,banerjee2016proximate,PhysRevB.86.144404,PhysRevB.84.024406,PhysRevB.90.195102,PhysRevB.85.134416,PhysRevB.87.195113,wang14}. Two paradigmatic examples are the Heisenberg antiferromagnet, where strong quantum fluctuations lead to a reduced magnetic moment in the ground state~\cite{low2009properties, jiang2012high, PhysRevB.49.3997}, and the Kitaev model, whose anisotropic bond-dependent interactions accommodates remarkable QSL phases~\cite{kitaev2006anyons, PhysRevLett.130.156701, PhysRevResearch.2.033318, PhysRevLett.123.087203, PhysRevB.102.121102, PhysRevB.108.085103}. These systems have attracted significant attention both for their fundamental theoretical importance and their experimental realization in materials such as $\alpha$-RuCl$_3$, where the competition between various (anisotropic and isotropic) exchange interactions, including Kitaev and Heisenberg terms, leads to rich physics~\cite{li2021identification,doi:10.1126/science.aah6015,trebst2017kitaevmaterials,lee20,sears2015magnetic}.

Tensor network methods offer powerful variational approaches for studying strongly correlated systems~\cite{PhysRevLett.69.2863,RevModPhys.77.259,schollwock2011density,Verstraete_2008,RevModPhys.93.045003, ORUS2014117,PhysRevB.88.165138,PhysRevB.95.024426,doi:10.1126/science.aam7127}. In particular, the infinite projected entangled-pair state (iPEPS) ansatz has emerged as an effective tool for representing 2D quantum states of infinite systems~\cite{verstraete2004renormalization, jordan2008classical,  PhysRevLett.112.147203, PhysRevB.81.165104,doi:10.1126/science.aam7127,PhysRevB.84.041108,nishio2004tensorproductvariationalformulation}. This approach combines computational efficiency with the ability to capture long-range quantum correlations, making it particularly suitable for systematically studying a wide range of models, including those on honeycomb lattices~\cite{jahromi2024kitaev,PhysRevB.107.054424,PhysRevB.108.085103,PhysRevB.100.035134,PhysRevX.2.041013,PhysRevB.111.L100402,gu13,gu20,nataf16}.


Accurate contraction of the 2D tensor network is required to evaluate observables and optimize iPEPS. One approach is to compute the effective environment tensors, which encapsulate the influence of the infinite system on a local region. The two most common techniques for this task are the corner transfer matrix renormalization group (CTMRG)~\cite{nishino1996corner,PhysRevB.80.094403,PhysRevB.98.235148} and the variational uniform matrix product state (VUMPS) method~\cite{PhysRevB.97.045145,10.21468/SciPostPhysLectNotes.7,PhysRevB.108.085103}. Both methods achieve controlled, low-rank approximations of the environment, albeit through different tensor representations. Such accurate environment tensors are fundamental for both the variational optimization of the iPEPS and the precise computation of its ground state properties.

The development of robust optimization algorithms leveraging these environments is crucial. Early strategies, such as the simple update (SU)~\cite{PhysRevLett.101.090603} and the more accurate but costly full update (FU)~\cite{ORUS2014117,jordan2008classical,PhysRevB.92.035142}, provided the first practical routes for ground state search. SU employs a local, approximate environment for efficiency, whereas FU utilizes a more complete environment approximation, typically computed via CTMRG, at a significantly higher computational cost. A transformative advancement has been the integration of automatic differentiation (AD)~\cite{ADTN} with CTMRG or VUMPS for gradient-based energy optimization~\cite{PhysRevB.94.035133,PhysRevB.94.155123}. This approach forms an efficient and powerful framework for variational ground state searches in 2D quantum many-body systems.

Meanwhile, recent developments in CTMRG have significantly expanded its applicability to a wider range of lattice geometries beyond square lattices. This includes various non-square lattices such as Bethe, hyperbolic, and honeycomb lattices~\cite{Daniška_2015,Daniška_2016,PhysRevE.78.061119,doi:10.1143/JPSJ.76.084004,PhysRevE.109.045305}. For the honeycomb lattice, specialized CTMRG approaches now enable efficient contraction while preserving the native $\C_{3v}$ symmetry, eliminating the need for artificially enlarged unit cells~\cite{PhysRevB.107.054424,PhysRevE.108.064132}. Parallel progress on square lattices has revealed that QR-based CTMRG implementations~\cite{qrctm}, which exploit $\C_{4v}$ symmetry, can achieve order-of-magnitude speedups compared to conventional SVD-based CTMRG~\cite{nishino1996corner}. Despite these advances, the efficient application of these accelerated contraction schemes to more general lattice geometries remains an open challenge.

In this work, we introduce a QR-based CTMRG framework tailored for iPEPS on honeycomb lattices, for highly efficient iPEPS contraction and optimization. This development achieves two key breakthroughs: First, it successfully extends the dramatic computational speedup of QR-based methods to the honeycomb geometry, enabling variational ground state searches with significantly reduced computational cost and the potential to explore higher precision regimes. Second, it demonstrates the applicability of QR-accelerated CTMRG algorithms beyond $\C_{4v}$ square lattices, providing a blueprint for extending these advantages to other lattice geometries.

We benchmark the performance and accuracy of our QR-based CTMRG framework against well-established methods, including quantum Monte Carlo (QMC)~\cite{low2009properties,jiang2012high} and conventional SVD-based CTMRG, for both the Heisenberg and Kitaev models on the honeycomb lattice. Our results demonstrate that the QR-based approach delivers significant computational speedups, which intrinsically links enhanced efficiency with the ability to explore larger bond dimensions ($D$ and/or $\chi$), thereby achieving comparable or superior accuracy in ground-state energies. For instance, by employing finite correlation length scaling (FCLS)~\cite{PhysRevB.99.245107,PhysRevX.8.031030,FCLS}, we obtain extrapolated values of the staggered magnetization and energy for the antiferromagnetic Heisenberg model that show excellent agreement with established QMC benchmarks.

Furthermore, we apply our QR-based CTMRG framework to analyze spatial correlation functions within the Kitaev-Heisenberg model~\cite{PhysRevLett.105.027204}. We specifically investigate the dimer-dimer correlation function deep within the spin-liquid regime~\cite{PhysRevResearch.6.033168,PhysRevB.95.024426,PhysRevB.106.174416,PhysRevLett.119.157203}, revealing clear numerical evidence of a universal algebraic decay proportional to $1/r^4$~\cite{PhysRevA.78.012304,PhysRevResearch.2.013005}. This theoretically predicted behavior for the pure Kitaev model is confirmed through large-scale iPEPS simulations, even in the presence of finite Heisenberg coupling.

The remainder of this paper is organized as follows. In Sec.~II, we introduce the technical details of our QR-based CTMRG approach tailored for honeycomb lattices. Sec.~III presents comprehensive benchmark results comparing our method's performance and accuracy against conventional approaches for both the antiferromagnetic Heisenberg model and the isotropic Kitaev model. We also analyze dimer-dimer correlation functions within the Kitaev-Heisenberg model, providing clear numerical evidence for the theoretically predicted universal $1/r^4$ decay characteristic of the QSL phase. Finally, Sec.~IV summarizes our principal findings and outlines promising directions for future research.

\section{Methods}
\subsection{iPEPS simulations on the honeycomb lattice\label{sec:ipeps}}

\begin{figure}[t]
	\centering
	\includegraphics[width=0.495\textwidth]{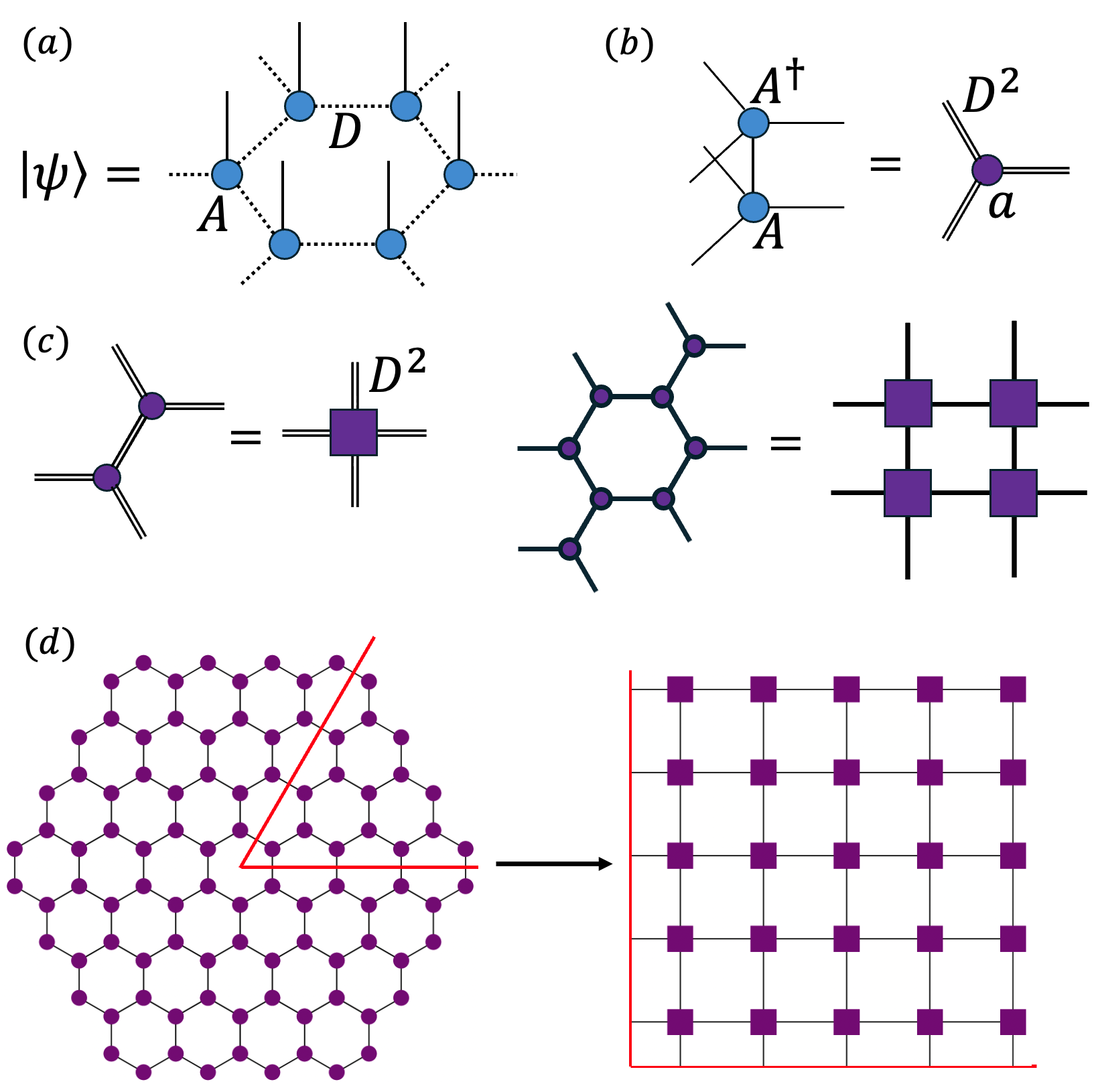}
	\caption{\label{fig:1} Schematic illustration of the iPEPS ansatz on the honeycomb lattice and the local mapping procedure for CTMRG. (a) iPEPS representation: blue circles denote local tensors $A$, with dashed lines representing virtual bonds of dimension $D$. (b) The double-layer tensor network, formed by contracting the iPEPS with its conjugate and full open lines, yields an effective tensor with a virtual bond dimension of $D^2$. (c) Local mapping procedure: two neighboring three-leg tensors (left) are mapped onto a single four-leg tensor (purple square), effectively transforming the hexagonal unit cell into a square one. (d) Global application of the local mapping procedure: a 60-degree angular sector (demarcated by red lines) of the honeycomb lattice is transformed into an effective square lattice patch, enabling the application of conventional CTMRG.}
\end{figure}

We employ the iPEPS ansatz to study quantum many-body systems on the honeycomb lattice. As shown in Fig.~\ref{fig:1}(a), the iPEPS wave function is represented by an infinite tensor network where each site is associated with a local tensor with bond dimension $D$. Evaluating physical observables and the state norm requires the contraction of the double-layer tensor network, depicted in Fig.~\ref{fig:1}(b).

A key challenge in applying the corner transfer matrix renormalization group (CTMRG) algorithm to non-square lattices, such as the honeycomb lattice, arises from the fact that conventional CTMRG is fundamentally designed for square or rectangular geometries. Historically, this limitation has been addressed by various strategies. One common approach involves defining a unit cell comprising two tensors, forming a four-leg building block that can be periodically tiled to construct an effective square lattice in Fig.~\ref{fig:1}(c), thereby restoring the periodicity required for standard CTMRG application~\cite{PhysRevX.2.041013}.

More recently, advanced approaches use the inherent rotational and mirror symmetries of iPEPS on the honeycomb lattice~\cite{PhysRevB.107.054424,PhysRevE.108.064132}. These methods focus on contracting only a 60-degree angular sector of the tensor network, with the full environment subsequently reconstructed by exploiting the lattice's rotational symmetry in Fig.~\ref{fig:1}(d). This sector-based construction is a direct generalization of the corner transfer matrix renormalization group idea of decomposing the environment into angular sectors~\cite{doi:10.1143/JPSJ.76.084004,PhysRevE.78.061119,Daniška_2015,Daniška_2016}, here adapted to the $\C_{3v}$ symmetry of the honeycomb lattice. In contrast to the previous approach, which simplifies the geometry at the cost of breaking rotational invariance, this approach preserves the native $\C_{3v}$ symmetry of the honeycomb lattice throughout the contraction.

The construction of the $\C_{3v}$-symmetric iPEPS tensor follows the approach in~\cite{PhysRevResearch.2.033318,PhysRevLett.123.087203,PhysRevB.107.054424} and is fully compatible with general spin-$S$ systems. The projection onto the $\C_{3v}$-symmetric subspace ensures that the double-layer tensor exhibits Hermiticity under reflection, as illustrated in Fig.~\ref{fig:2}(b). This property is necessary for the QR-based CTMRG, as it guarantees the Hermiticity of the enlarged corner tensors. All technical details, including the explicit unitary and anti-unitary transformations and their role in the algorithm for general spin-$S$, are provided in Appendix~\ref{app:c3v}.

\subsection{CTMRG for honeycomb lattice: SVD-based and QR-based approaches}

\begin{figure}[t]
	\centering
	\includegraphics[width=0.495\textwidth]{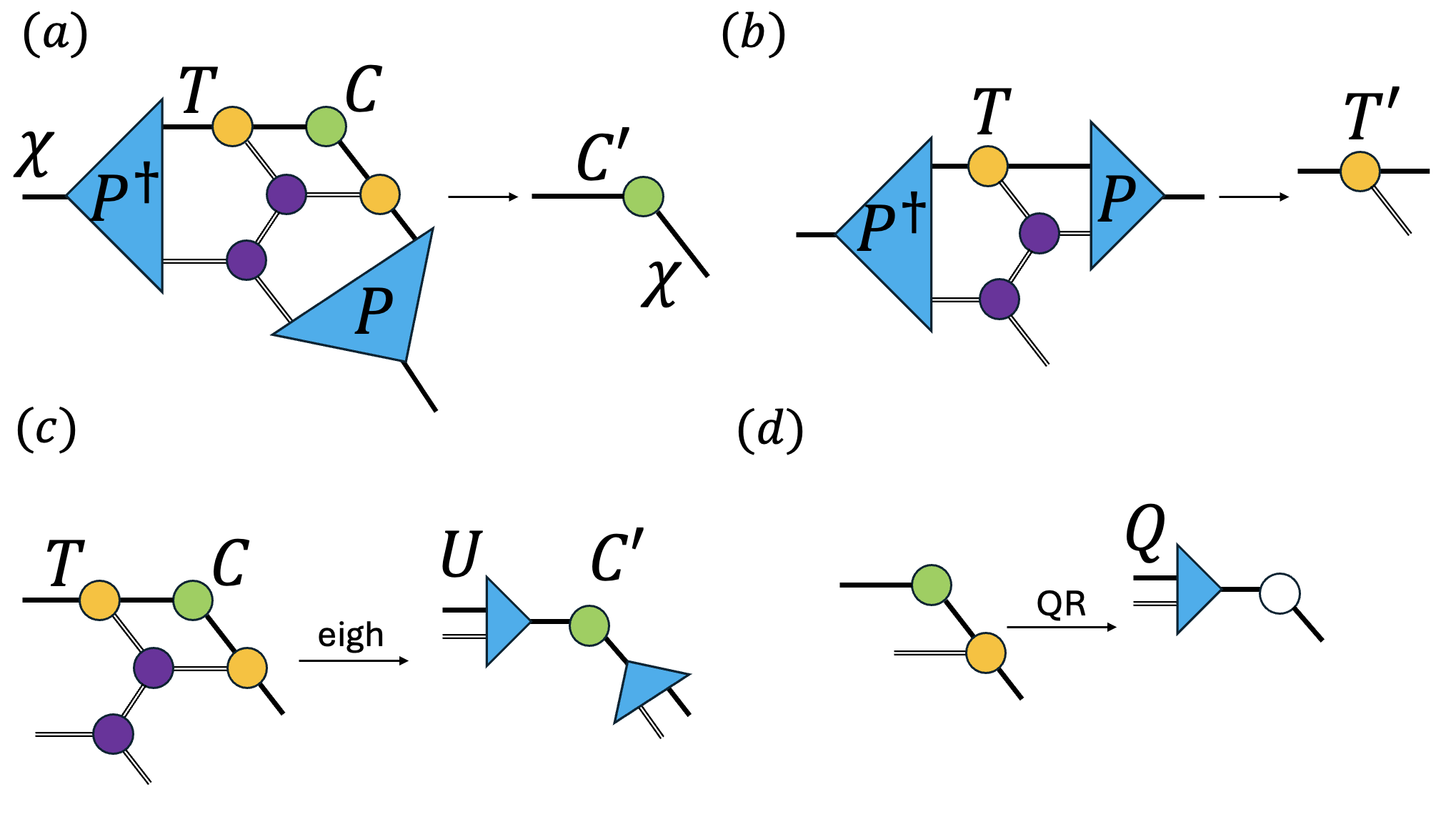}
	\caption{\label{fig:3} Schematic illustration of the CTMRG algorithm on the honeycomb lattice. (a) Corner tensor update rule: the updated corner tensor $C'$ is obtained by contracting the existing corner $C$ and edge $T$ tensors with the isometric projectors $P$ and $P^\dagger$. (b) Edge tensor update rule: the updated edge tensor $T'$ is obtained by contracting the existing edge $T$ tensor with the projectors $P$ and $P^\dagger$. (c) Projector construction for conventional SVD/EVD-based CTMRG: an enlarged corner tensor (formed by contracting network segments involving $T$ and $C$) undergoes eigenvalue decomposition (denoted ``eigh'') to obtain the projector $U$ (serving as $P$) for bond dimension truncation. (d) Projector construction for QR-based CTMRG: reduced corner tensor undergoes QR decomposition, yielding a unitary matrix $Q$ (serving as $P$) and an upper triangular matrix (white circle).}
\end{figure}

The CTMRG algorithm iteratively refines the environment tensors—corner tensors ($C$) and edge tensors ($T$)—that approximate the infinite tensor network surrounding a central unit cell. In the honeycomb lattice CTMRG, these updates are performed on the effective tensors derived from either the double-layer construction or within a 60-degree angular sector, as discussed in the previous Sec.~\ref{sec:ipeps}~\cite{PhysRevB.107.054424}. The environment is grown iteratively: starting from small initial corner and edge tensors, the system size is gradually increased while applying the isometry $P$ (and its conjugate $P^\dagger$) to truncate virtual bond dimensions up to $\chi$ as illustrated in Fig.~\ref{fig:3}(a), retaining the most significant contributions to the environment.

The general update rules for the corner and edge tensors are depicted in Fig.~\ref{fig:3}(a) and Fig.~\ref{fig:3}(b), respectively. Specifically, once an isometric projector $P$ is obtained, the updated corner tensor $C'$ is constructed by contracting the current corner $C$ and edge $T$ tensors with the projectors $P$ and $P^\dagger$ [Fig.~\ref{fig:3}(a)]. Similarly, the updated edge tensor $T'$ is obtained by contracting the current edge $T$ tensor with the projectors $P$ and $P^\dagger$ [Fig.~\ref{fig:3}(b)]. This iterative process continues until the environment tensors converge to within a desired precision threshold.

Within the conventional symmetric CTMRG (enlarged corner tensors are Hermitian) framework, our approach differs in how the projectors are obtained. In the standard approach, the projector is computed via singular value decomposition (SVD) or, more efficiently, via eigenvalue decomposition (EVD) of the enlarged corner tensor, as illustrated in Fig.~\ref{fig:3}(c). The virtual bond dimension is then truncated by discarding components with small singular values or small-magnitude eigenvalues.

Our work employs a QR-based CTMRG framework tailored for the honeycomb lattice, drawing inspiration from recent advancements in QR-based tensor network methods for the square lattice~\cite{qrctm}. These approaches have demonstrated remarkable effectiveness, achieving accuracy comparable to conventional methods while offering superior computational performance.

As illustrated in Fig.~\ref{fig:3}(d), rather than performing an eigenvalue decomposition on an enlarged corner tensor, our approach directly subjects a specific tensor network segment to QR decomposition. This decomposition yields an unitary matrix $Q$ and an upper triangular matrix $R$. The unitary matrix $Q$ directly serves as the isometric projector $P$ for truncating the virtual bond dimensions. This approach effectively provides a valid projector for dimension reduction while avoiding the computationally intensive explicit construction and diagonalization of the enlarged corner tensor required in conventional methods.

This QR-based scheme significantly enhances computational efficiency, particularly for large bond dimensions, due to two key advantages: (1) Algorithmic efficiency: QR decomposition algorithms offer superior computational efficiency compared to EVD or SVD, being highly amenable to parallelization and optimized for modern GPU architectures. (2) Reduced problem size: The QR decomposition operates on a tensor reshaped into a matrix of dimensions $\chi D \times \chi$, whereas conventional SVD/EVD-based CTMRG typically requires decomposition of a matrix with dimensions $\chi D \times \chi D$. This reduction in problem size drastically lowers computational complexity and memory requirements, enhancing scalability for larger bond dimensions.

\subsection{Measurements}

\begin{figure}[t]
	\centering
	\includegraphics[width=0.495\textwidth]{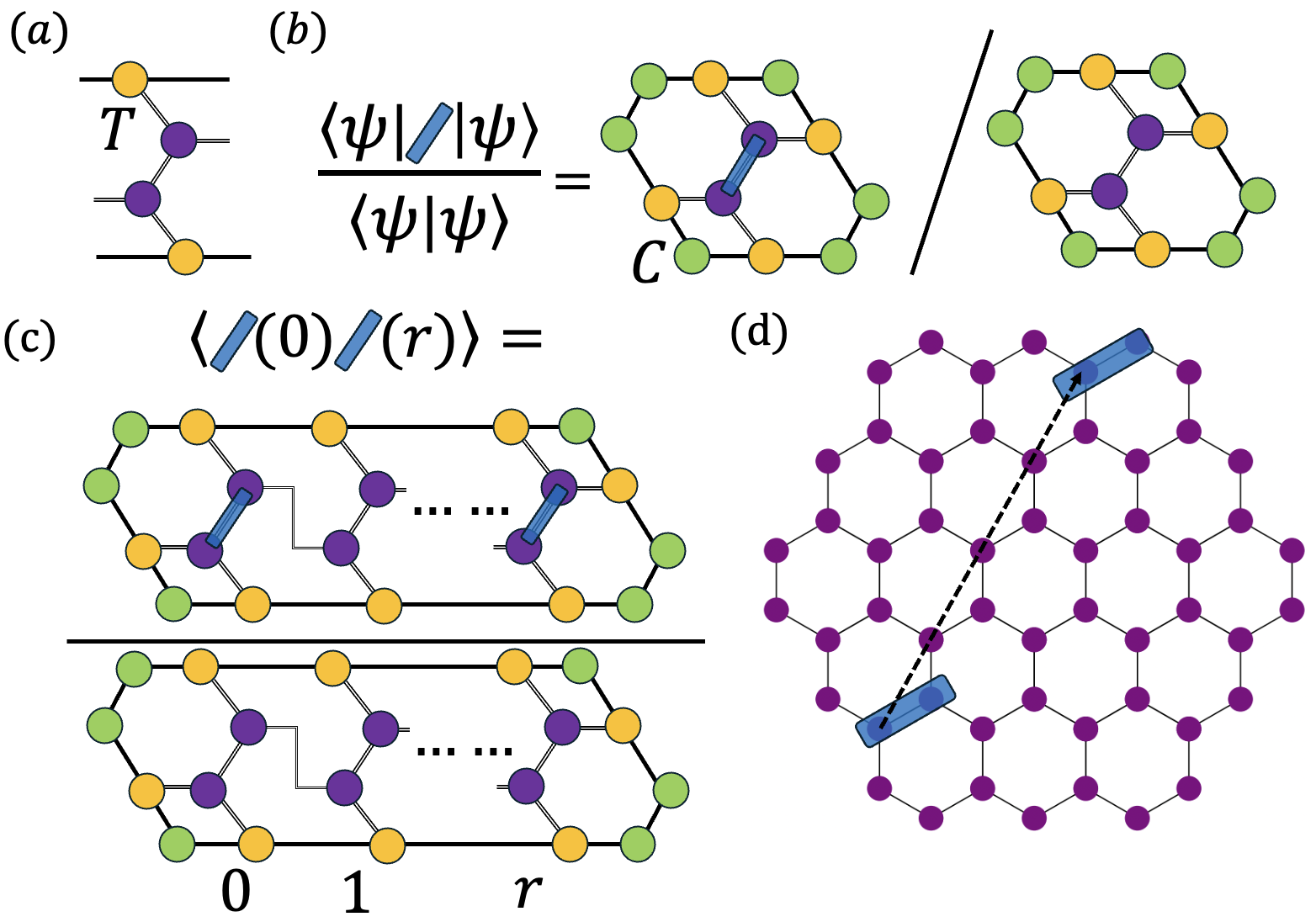}
	\caption{Schematic illustration of physical observable calculations in iPEPS on the honeycomb lattice. (a)~Construction of the effective transfer matrix from the converged environment. (b)~Calculation of local expectation values by contracting the operator with the iPEPS tensor and its environment. (c)~Contraction scheme for two-point correlation functions $\langle O(0) O(r) \rangle$ between local operators $O(0)$ and $O(r)$. (d)~In this work, we adopt this relative spatial arrangement of two dimer operators on the honeycomb lattice.}
	\label{fig:measurements}
\end{figure}

After the iPEPS tensor $A$ and its corresponding CTMRG environment tensors $(C,T)$ have converged, local observables can be efficiently computed by contracting the network illustrated in Fig.~\ref{fig:measurements}(b), representing the formula:
\begin{equation}
\langle O \rangle = \frac{\bra{\Psi}O\ket{\Psi}}{\braket{\Psi|\Psi}}.
\end{equation}

For dimer-dimer correlations along the direction specified in Fig.~\ref{fig:measurements}(d), we compute the connected correlation function
\begin{equation}
C_{OO}(r) = \langle O(0) O(r) \rangle - \langle O \rangle^2,
\label{eq:Coo}
\end{equation}
as represented in Fig.~\ref{fig:measurements}(c).


The correlation length $\xi$ can be determined using two complementary approaches. First, one can compute it from the eigenvalue of the transfer matrices shown in Fig.~\ref{fig:measurements}(a)~\cite{nishino1996corner}. We define $\lambda_i'$ as the i-th-largest magnitude of the set of eigenvalues and its normalized magnitude as $\lambda_i=\lambda_i'/\lambda_0'$. The correlation length associated with the slowest-decaying mode is then
\begin{equation}
\xi = -\frac{1}{\ln \lambda_1}.
\end{equation}
Alternatively, one can perform a numerical fit to the decay of the measured $C_{OO}(r)$ at sufficiently large distances to estimate the correlation length associated with operator $O$. This real-space fitting not only provides a consistency check for the transfer matrix approach but is also essential if the physically relevant mode is not the slowest-decaying one in the system~\cite{PhysRevLett.129.177201}.

We employ Finite Correlation Length Scaling (FCLS) to extrapolate our numerical results to the infinite correlation length limit for gapless systems~\cite{PhysRevX.8.031030,FCLS}. This technique is applicable to the case where the finite bond dimension $D$ induces a finite correlation length $\xi$, which acts as a cutoff to the diverging correlation length, analogous to a finite system size in conventional finite-size scaling, such as Lorentz-invariant critical points~\cite{PhysRevX.8.031030,FCLS}. This technique utilizes the systematic dependence of physical observables on the finite correlation length $\xi$ inherent in simulations with bounded bond dimensions. Specifically, for a fixed iPEPS bond dimension $D$, we perform CTMRG measurements at various environment bond dimensions $\chi$ to obtain physical observables $O(D,\chi)$ and the few leading eigenvalues of the CTMRG transfer matrix.

To accurately determine the correlation length $\xi(D)$ for a given $D$ in the infinite $\chi$ limit, we first need to eliminate the truncation effects of the CTMRG virtual bond dimension $\chi$. This is achieved by extrapolating the spectral gap of the transfer matrices to the $\chi \rightarrow \infty$ limit. More precisely, for each $D$, we extrapolate $\xi(D,\chi\rightarrow\infty)$ with the formula~\cite{PhysRevX.8.041033}:

\begin{equation}
	\xi^{-1}(D,\chi) = \xi^{-1}(D,\infty) + a \ln\left(\frac{\lambda_1(D,\chi)}{\lambda_2(D,\chi)}\right).
	\label{eq:fcls}
\end{equation}

For each bond dimension $D$, we take the observable $O(D,\chi_{\text{max}})$ from the largest available $\chi$ and pair it with the corresponding extrapolated correlation length $\xi(D,\chi\rightarrow \infty)$. The $(O,\xi)$ pairs provide a convenient representation of the convergence of the observable with correlation length. By extrapolating $O(\xi)$ to the $\xi \rightarrow \infty$ limit, we obtain accurate estimates for the physical quantities in the limit of infinite $D$ and $\chi$.

\subsection{Variational optimization of iPEPS with automatic differentiation}

A cornerstone of our optimization framework is the integration of automatic differentiation (AD) for computing energy gradients~\cite{ADTN}. The application of AD to iPEPS optimization represents a well-established and robust technique that offers several significant advantages. First, AD provides exact analytical gradients, eliminating the numerical approximations inherent in finite-difference methods. Second, it dramatically simplifies the implementation of gradient calculations, as derivatives are automatically propagated through the entire tensor network contraction process. This capability enables a clean and efficient code structure, which is particularly beneficial for challenging contraction algorithms. Building on its successful application in QR-based CTMRG for square lattices with $\C_{4v}$ symmetry, AD integrates with our CTMRG framework on the honeycomb lattice. We employ \textit{PyTorch} as the AD backend~\cite{pytorch}. A demonstration of our method is publicly available~\cite{GithubQRCTM}.

To determine the ground state of the system, we employ a variational optimization scheme for the iPEPS tensor $A$ that minimizes the energy expectation value computed using the tensor network contraction illustrated in Fig.~\ref{fig:measurements}(b), where the operator $O$ corresponds to the local Hamiltonian term. This minimization is performed using the L-BFGS optimizer implemented in \textit{PyTorch}. The optimization for a given bond dimension $D$ is typically initiated from a previously converged state at a lower bond dimension, whose iPEPS tensor $A$ is padded with a small random perturbation to fit the new virtual dimension.

\section{Results}

\subsection{Kitaev-Heisenberg Hamiltonian}

To demonstrate the effectiveness of our method, we focus on the Kitaev-Heisenberg (KH) model, described by the Hamiltonian
\begin{equation}
	H = \sum_{\langle ij \rangle_\alpha} K S_i^\alpha S_j^\alpha + J \sum_{\langle ij \rangle} \vec{S}_i \cdot \vec{S}_j,
\end{equation}
where the bond-dependent Kitaev interactions of strength $K$ act on the three distinct types of nearest-neighbor bonds $\langle ij \rangle_\alpha$ (with $\alpha = x,y,z$), reflecting the underlying honeycomb lattice symmetry, and the Heisenberg interaction of strength $J$ acts isotropically on all nearest-neighbor bonds.

\subsection{Antiferromagnetic Heisenberg model on the honeycomb lattice}

To assess the computational efficiency and numerical accuracy of our approach, we first benchmark our method in the spin-1/2 antiferromagnetic Heisenberg limit ($J=1, K=0$). This model provides an ideal testbed due to the availability of reliable Quantum Monte Carlo (QMC) results for accuracy validation~\cite{low2009properties,jiang2012high}. To enable the use of a uniform iPEPS ansatz for this antiferromagnet, we apply a $\pi$-rotation about the $y$-axis on every site of the B sublattice, which maps the staggered magnetic order onto a uniform configuration~\cite{hasik2021investigation,PhysRevB.96.121118,PhysRevLett.133.176502}. We structure our benchmarks into two complementary parts: (i) computational efficiency tests, featuring a direct performance comparison between QR-based and SVD-based CTMRG in computational speed, numerical accuracy, and a similar convergence behavior, and (ii) a high-precision comparison between our FCLS-extrapolated results and established QMC values derived from finite-size scaling.

\begin{figure}[t]
	\includegraphics[width=\columnwidth]{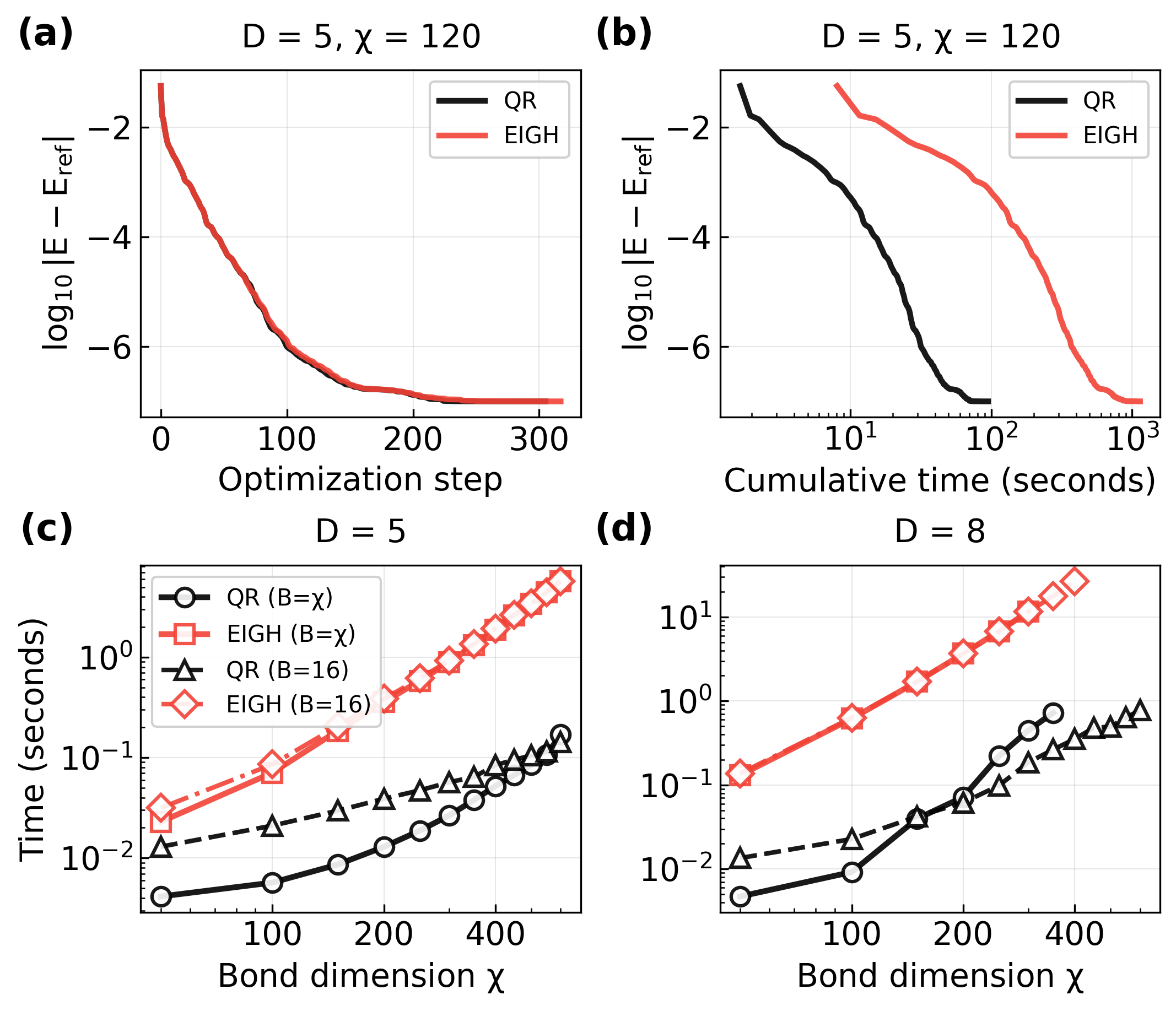}
	\caption{Computational efficiency benchmark comparing QR-based (Black) and Eigh-based (Red) CTMRG for the spin-1/2 antiferromagnetic Heisenberg model ($K=0, J=1$) on a single NVIDIA H100 80G GPU. (a)~Convergence of variational energy per site as a function of optimization steps. Bond dimension $D=5$ and unsliced environment bond dimension $\chi=120$ with $k=24$ CTMRG steps for gradient tracking and reference energy $E_{\text{ref}} = -0.5444744$. (b)~Same is in~(a)~but plotted as a function of the cumulative wall time.  (c)~Average runtime per CTMRG forward iteration as a function of $\chi$ for $D=5$. Circles (QR-based, unsliced) and squares (Eigh-based, unsliced) indicate unsliced contraction ($B=\chi$) while triangles (QR-based, sliced) and diamonds (Eigh-based, sliced) indicate fixed batch size ($B=16$). The QR approach consistently achieves equivalent accuracy at significantly reduced computational cost. Missing data points indicate configurations where the Eigh-based method encountered memory limitations. (d)~Same is in (c) but $D=8$.}
	\label{fig:benchmark}
\end{figure}

In our implementation, we adopt a slicing strategy to avoid storing large intermediate tensors during the computation of tensor network segments shown in Fig.~\ref{fig:3} and Fig.~\ref{fig:measurements}. This is achieved by slicing the virtual bond dimension $\chi$ along a single open leg into sub-tensors with bond dimension $B$ along the sliced leg. Each sub-tensor is contracted independently with the remaining part of the tensor network segment and the resulting partial tensors are reassembled into the final tensor. The value of $B$ is chosen to balance memory usage and parallel efficiency.

As a first step, we compare the computational efficiency of QR-based and Eigh-based CTMRG for moderate parameters on a single NVIDIA H100 80G GPU. Fig.~\ref{fig:benchmark} summarizes these results. For bond dimension $D=5$ and environment bond dimension $\chi=120=B$, we track $k=24$ CTMRG steps for gradient collection, preceded by $10$ warm-up CTMRG steps. Both approaches converge to the same final energy and a similar convergence behavior as shown in Fig.~\ref{fig:benchmark}(a) and Fig.~\ref{fig:benchmark}(b). The advantage of the QR-based implementation becomes evident: achieving the same energy accuracy takes only 95 seconds with QR, compared to 1136 seconds for the Eigh-based method.

The lower panels of Fig.~\ref{fig:benchmark} further quantify the computational cost per CTMRG iteration for bond dimensions $D=5$ and $D=8$ across a range of environment bond dimensions $\chi$. In practice, the QR-based approach is significantly faster across the entire range of $\chi$, achieving speedups of up to 75$\times$ at parameter values $(D,\chi,B)=(8,400,16)$. This performance gain stems from a drastically reduced prefactor in its computational scaling, while the leading-order complexity is the same as the Eigh-based method. Notably, the Eigh-based method encountered memory limitations for large bond dimensions, which is the reason for the missing data points at large $\chi$ in Figs.~\ref{fig:benchmark}(d). This also highlights a significant advantage of the QR-based approach in terms of memory footprint and scalability. Slicing strategies also play a crucial role in these benchmarks: employing a fixed batch size of $B=16$ provides additional performance improvements over unsliced contraction ($B=\chi$) at larger $\chi$ values, suggesting a more optimized utilization of GPU resources when processing smaller, fixed-size batches for this task.

We obtained optimized iPEPS tensors for bond dimensions ranging from $D=3$ to $8$, with corresponding environment bond dimensions during optimization of $\chi_{\text{opt}}=100$ to $350$. Physical observables were calculated with environment bond dimensions up to $\chi=700$ for a higher precision in the final evaluation.

\begin{table}[htbp]
	\centering
	\begin{adjustbox}{max width=\textwidth}
		\begin{tabular}{
			r
			r
			r
			c
			r
			S[table-format=-1.8]
			S[table-format=1.4]
			S
			}
			\toprule
			{$D$}            & {$\chi_{\text{opt}}$} & {$k$} & {$\mathrm{U}(1)$ charges}                   & {$\chi_{\text{max}}$} & {$E(\chi_{\text{max}})$} & {$m^2(\chi_{\text{max}})$} & {$\xi_T/\xi_L$} \\
			\midrule
			$\boldsymbol{3}$ & 100                   & 16    & $(002)$                            & 600                   & -0.53986093              & 0.1194                     &                 \\
			$\boldsymbol{4}$ & 150                   & 20    & $(\overline{2}002)$                & 600                   & -0.54393504              & 0.0948                     & 2.0             \\
			$\boldsymbol{5}$ & 200                   & 24    & $(\overline{2}0022)$               & 600                   & -0.54447430              & 0.0840                     & 2.0             \\
			$\boldsymbol{6}$ & 250                   & 28    & $(\overline{2}00022)$              & 600                   & -0.54449300              & 0.0831                     & 2.0             \\
			$6$              & 250                   & 28    & $(\overline{2}\overline{2}0022)$   & 700                   & -0.54454394              & 0.0801                     & 2.1             \\
			$\boldsymbol{7}$ & 300                   & 32    & $(\overline{2}000222)$             & 700                   & -0.54451062              & 0.0825                     & 2.0             \\
			$7$              & 300                   & 32    & $(\overline{2}\overline{2}00022)$  & 700                   & -0.54456716              & 0.0793                     & 1.5             \\
			$\boldsymbol{7}$ & 300                   & 32    & $(\overline{2}\overline{2}00224)$  & 700                   & -0.54456196              & 0.0796                     & 2.1             \\
			$8$              & 350                   & 36    & $(\overline{2}\overline{2}000022)$ & 500                   & -0.54457409              & 0.0786                     & 1.5             \\
			$\boldsymbol{8}$ & 350                   & 36    & $(\overline{2}\overline{2}000222)$ & 500                   & -0.54458031              & 0.0781                     & 2.1             \\
			\bottomrule
		\end{tabular}
	\end{adjustbox}
	\caption{Optimization parameters and results for the spin-1/2 antiferromagnetic Heisenberg model on the honeycomb lattice. $D$: bond dimension; $\chi_{\text{opt}}$: environment bond dimension used during AD optimization; $k$: number of CTMRG steps tracked for gradients; $\mathrm{U}(1)$ charges: symmetry sector configuration, $\overline{2} \equiv -2$; $\chi_{\text{max}}$: maximum environment bond dimension used for measurements; $E$: energy per site; $m^2$: square of staggered magnetization per site; $\xi_T$($\xi_L)$ fitted transverse (longitudinal) correlation length. The bold number $D$ data are used for FCLS in Fig \ref{fig:heisenberg}.\label{tab:heisenberg_data}}
\end{table}

\begin{figure}[t]
	\centering
	\includegraphics[width=0.495\textwidth]{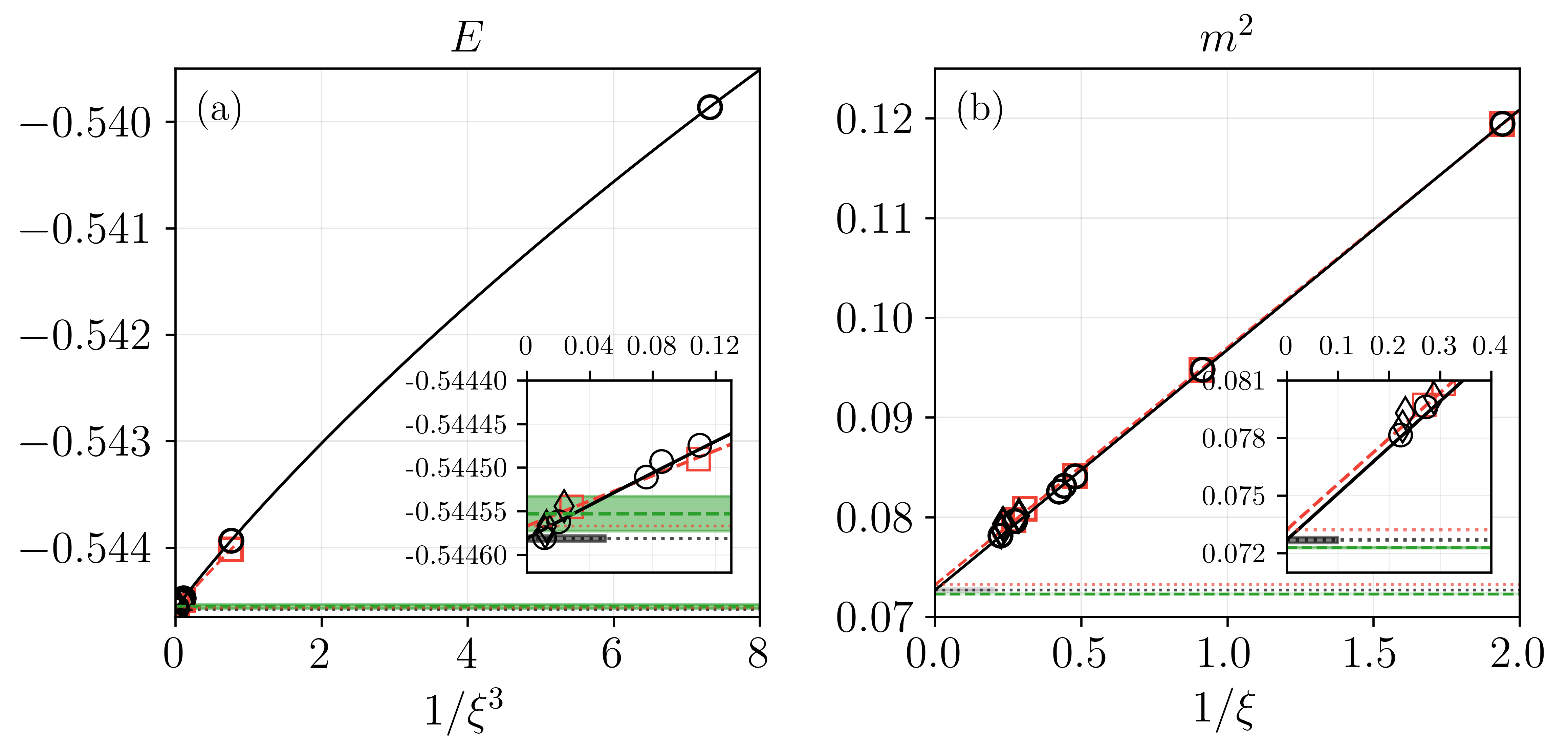}
	\caption{FCLS for the spin-1/2 antiferromagnetic Heisenberg model on the honeycomb lattice. (a) Ground state energy per site $E$ as a function of $1/\xi^3$. (b) Squared staggered magnetization per site $m^2$ as a function of $1/\xi$. Black circles represent our data points used for FCLS fits, with solid lines and error bands indicating the corresponding fits. Black diamonds represent data points not included in the fits. Red squares and dashed red lines show reference SVD-based data without $\mathrm{U}(1)$ symmetry from~\cite{PhysRevB.107.054424}. Green bands with error bars indicate QMC results from~\cite{low2009properties,jiang2012high}. The inset in (a,b) provides a magnified view of the convergence of energy and staggered magnetization near the infinite $\xi$ limit.}
	\label{fig:heisenberg}
\end{figure}

To achieve more accurate correlation length scaling, we enforced $\mathrm{U}(1)$ symmetry on the iPEPS tensors. This was achieved by setting elements that do not satisfy the $\mathrm{U}(1)$ charge conservation rules in the dense iPEPS tensor $A$ to zero. Specifically, the sum of charges from all legs of $A$ must equal $1$. On the physical leg, the $\mathrm{U}(1)$ charges are $\pm 1$ and the virtual-leg charges along with other detailed specifications and results are provided in TABLE.~\ref{tab:heisenberg_data}.

We employ FCLS to extrapolate our results in TABLE.~\ref{tab:heisenberg_data} to the infinite $\xi$ limit. Only data points marked with bold $D$ are used for extrapolation, as they constitute a consistent family of states converging to the critical point, which is a necessary condition for valid FCLS~\cite{hasik2021investigation}. Other points exhibit different scaling behaviors or non-negligible offsets, which we attribute to them not belonging to the same family of states, potentially due to specific symmetry sector choices at higher bond dimensions. We measured the transverse correlation length $\xi_T$ and longitudinal correlation length $\xi_L$. The anisotropy ratio $\xi_T/\xi_L$, as observed for $D=3$-$8$, is consistent with the established universal scaling behavior for 2D quantum antiferromagnets on the square lattice~\cite{hasik2021investigation}.

For the ground state energy per site $E$, we use the scaling form $E(\xi) = E_0 + a/\xi^3 + b/\xi^4$. As shown in Fig.~\ref{fig:heisenberg}(a), our calculated energies for $D=3$-$8$ exhibit excellent agreement with this scaling form when plotted against $1/\xi^3$ (with a higher-order correction $b/\xi^4$ included in the fit). The extrapolated energy from these data points yields $E_0 = -0.544582(3)$, showing remarkable agreement with the highly accurate Quantum Monte Carlo (QMC) benchmark $E_{\text{QMC}} = -0.544553(20)$~\cite{low2009properties}. Our variational energy at $E(D=8,\chi=500) = -0.54458031$ is already slightly lower than the QMC reference value, highlighting the high precision achieved.

For the squared staggered magnetization per site $m^2$, we use the scaling form $m^2(\xi)=m_0^2 + a/\xi$. As shown in Fig.~\ref{fig:heisenberg}(b), our iPEPS results for $D=3$-$8$ follow this linear scaling, enabling robust extrapolation to the infinite $\xi$ limit. The linear extrapolated staggered magnetization per site $m_0 = 0.2696(3)$ compares favorably with the QMC reference value $m_{\text{QMC}}=0.26885(2)$~\cite{jiang2012high} and is better than the FCLS extrapolated $m_{\text{SVD}}=0.2705(25)$ from~\cite{PhysRevB.107.054424}. Our results demonstrate that the QR-based CTMRG method achieves high precision on the honeycomb lattice, yielding ground state properties in excellent quantitative agreement with state-of-the-art numerical benchmarks.

\subsection{Isotropic Kitaev model}

The ground state energy of the spin-1/2 isotropic Kitaev model provides a useful benchmark due to its exact solvability, offering an ideal testbed for validating the accuracy and convergence properties of our QR-based CTMRG framework on the honeycomb lattice. We consider the ferromagnetic case ($K = -1$) in the following analysis. Although the Kitaev Hamiltonian is not strictly $\C_{3v}$-symmetric due to bond-dependent interactions, it has been shown that a $\C_{3v}$-symmetric CTMRG scheme can be applied as shown in Ref.~\cite{PhysRevB.107.054424} (see Appendix.~\ref{app:c3v} for the implementation guide for general spin-$S$). In this case, the representation of the $\C_{3v}$ in Eq.(\ref{eq:Uc}) and Eq.(\ref{eq:Up}) is complex-valued. Consequently, use of complex-valued tensors in our simulation is the most natural approach.

\begin{figure}[t]
	\includegraphics[width=\columnwidth]{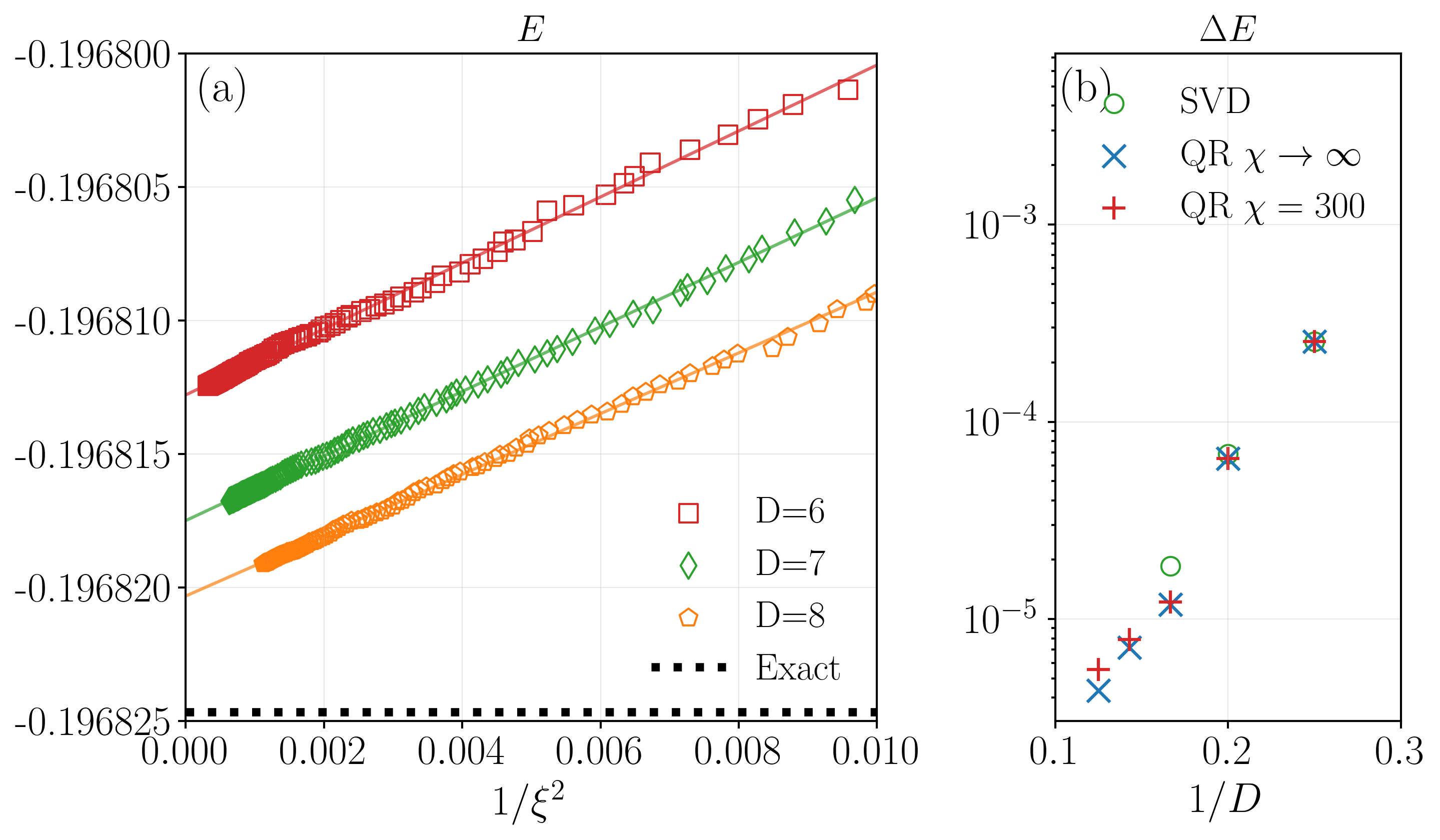}
	\caption{(a) Ground state energy per site $E$ as a function of $1/\xi^2$ for bond dimensions $D=6, 7, 8$. Solid lines represent linear extrapolations to the infinite correlation length limit ($1/\xi^2 \to 0$), yielding extrapolated energies $E$. The black dashed line indicates the exact ground state energy $E_{\text{exact}} = -0.19682465$. Vertical offsets are for visual clarity. (b) Absolute error $\Delta E = E - E_{\text{exact}}$ as a function of $1/D$ on a semi-logarithmic scale, demonstrating systematic convergence toward the exact solution. Green circles represent data from SVD-based approaches~\cite{PhysRevB.107.054424}, while plus (crosses) denote raw (extrapolated) results from our QR-based method.
	}\label{fig:kitaev_e}
\end{figure}

We optimized iPEPS tensors for bond dimensions $D=4$ to $8$ using environment bond dimensions $\chi_{\text{opt}}=80$ to $240$ during optimization, and computed physical observables with $\chi$ up to $300$. Other detailed specifications and results are provided in TABLE.~\ref{tab:kitaev_data}. The isotropic Kitaev model exhibits more pronounced finite-$\chi$ effects compared to the Heisenberg model, necessitating careful extrapolation.

Empirically, we observe $E(D,\xi(D,\chi)) \sim \xi^{-2}(D,\chi)$ scaling for all $D>5$, varying $\chi$, as shown in Fig.~\ref{fig:kitaev_e}(a). Our results demonstrate systematic convergence of $E$ toward the exact ground state energy $E_{\text{exact}} = -0.19682465$ with increasing bond dimensions. The quantitative precision is further illustrated in Fig.~\ref{fig:kitaev_e}(b), where the absolute error $\Delta E$ versus $1/D$ on a semi-logarithmic scale shows systematic decrease from $D=4$ to $8$, confirming the high accuracy of our approach. Our QR-based results achieves energies that are lower than or comparable to the SVD-based benchmarks reported in Ref.~\cite{PhysRevB.107.054424}. We note that a direct comparison of the methods is hindered by the lack of detailed environment bond dimension $\chi$ data in the prior work.

\begin{table}[htbp]
	\centering
	\begin{adjustbox}{max width=\textwidth}
		\begin{tabular}{
			r
			r
			r
			S[table-format=1.1e-1]
			S[table-format=-1.8]
			S[table-format=-1.8]
			}
			\toprule
			$D$ & $\chi_{\text{opt}}$ & $k$ & {$m$}  & {$E(\chi=300)$} & {$E(\chi\rightarrow\infty)$} \\
			\midrule
			4   & 80                  & 20  & 1.3e-5 & -0.19656874     & -0.19656874                  \\
			5   & 120                 & 24  & 2.4e-6 & -0.19675938     & -0.19675967                  \\
			6   & 160                 & 28  & 3.8e-6 & -0.19681244     & -0.19681278                  \\
			7   & 200                 & 32  & 8.2e-6 & -0.19681676     & -0.19681749                  \\
			8   & 240                 & 36  & 1.2e-6 & -0.19681910     & -0.19682031                  \\
			\bottomrule
		\end{tabular}
	\end{adjustbox}
	\caption{Optimization parameters and results for the spin-1/2 isotropic Kitaev model ($K=-1$). $D$: bond dimension; $\chi_{\text{opt}}$: environment bond dimension used during AD optimization; $k$: number of CTMRG steps tracked for gradients; $m$: the magnetization per site; $E$: energy per site.\label{tab:kitaev_data}}
\end{table}

\begin{figure}[t]
	\includegraphics[width=\columnwidth]{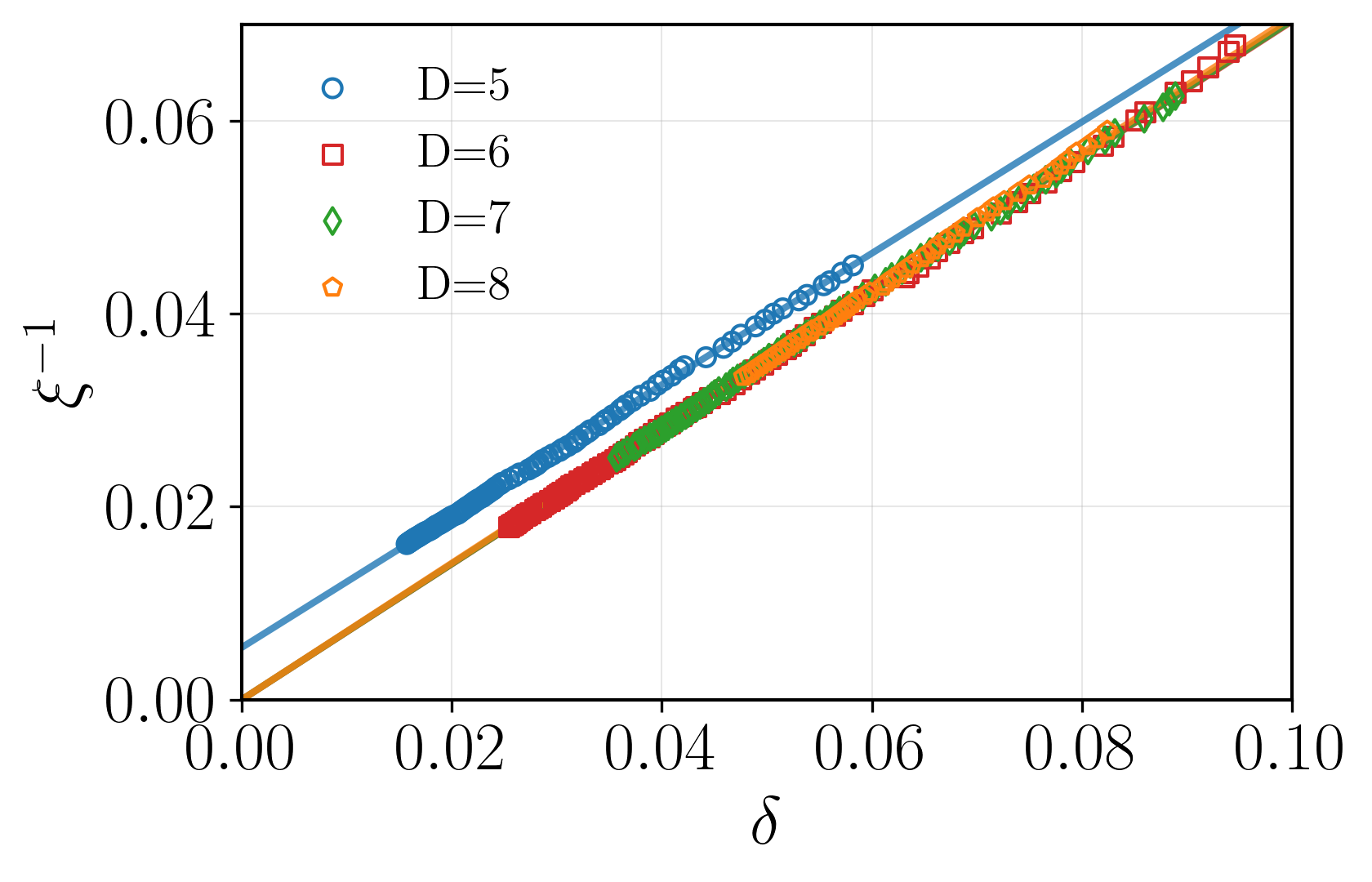}
	\caption{Scaling analysis for the isotropic Kitaev model. The inverse correlation length $\xi^{-1}$ is plotted against $\delta = \ln(\lambda_1/\lambda_2)$ according to Eq.(\ref{eq:fcls}) for bond dimensions $D=5,6,7,8$. Solid lines indicate linear fits. The data collapse demonstrates that while the iPEPS with $D=5$ remains gapped even in the infinite $\chi$ limit, the iPEPS with $D\geq6$ are gapless.\label{fig:fcsl}}
\end{figure}

Through high-precision simulations, we explicitly reveal the gapless nature of the isotropic Kitaev model within the iPEPS framework. Fig.~\ref{fig:fcsl} reveals a qualitative transition: while the $D=5$ ansatz remains gapped in the infinite $\chi$ limit, the $D\geq6$ simulations are clearly gapless, consistent with the known gapless property of the exact solution.

\subsection{Dimer-Dimer correlation in the Kitaev-Heisenberg model}

An open question in the Kitaev-Heisenberg model is whether the dimer-dimer correlations, rigorously established to follow a $1/r^4$ algebraic decay in the pure Kitaev spin liquid~\cite{PhysRevA.78.012304,PhysRevResearch.2.013005}, remain intact once Heisenberg interactions break integrability but the ground state is still within the Kitaev spin-liquid regime. Deep inside this regime, extensive evidence indicates that the system remains a gapless QSL with Dirac cones~\cite{PhysRevB.108.085103,PhysRevB.95.024426,PhysRevB.90.195102,PhysRevLett.105.027204,PhysRevResearch.6.033168,PhysRevLett.119.157203,PhysRevB.106.174416}, suggesting that the same scaling law for dimer-dimer correlations should hold. Yet, a direct numerical demonstration of the $1/r^4$ decay in the Kitaev-Heisenberg model has so far been missing.

Here, to close this gap, we compute the dimer-dimer correlation in both the pure Kitaev and Kitaev-Heisenberg models. The correlation is defined as
\begin{equation}
C_{D^zD^z}(r) = \langle D^z(0) D^z(r) \rangle - \langle D^z(0) \rangle \langle D^z(r) \rangle,
\end{equation}
where $D^z(r) = S^z_i(r) S^z_j(r)$ denotes the $z$-dimer operator on the $z$-bond $\langle ij\rangle$ at position $r$, as illustrated in Fig.~\ref{fig:measurements}(c).

\begin{figure}[t]
	\includegraphics[width=\columnwidth]{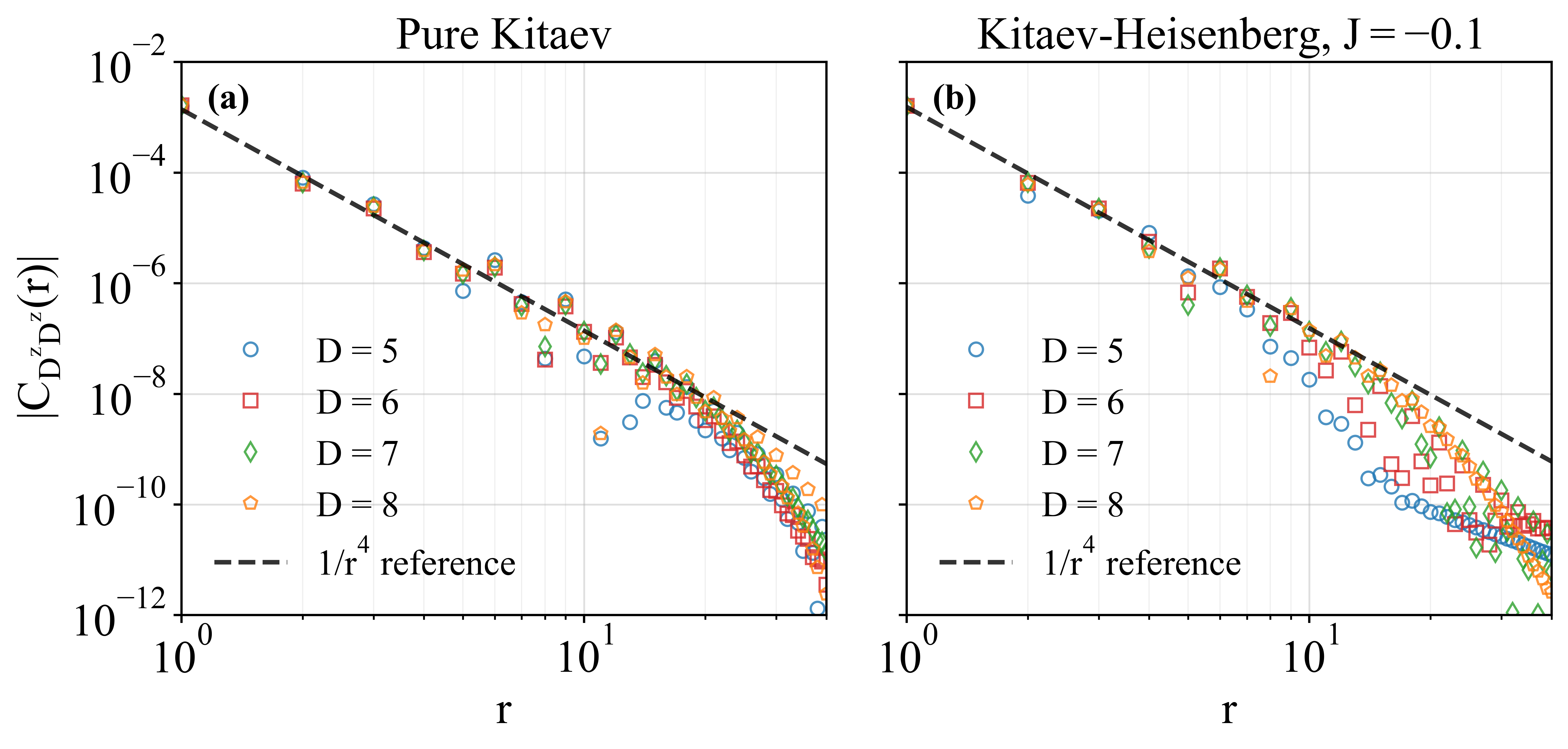}
	\caption{\label{fig:dimerdimer} Absolute values of dimer-dimer correlation function $C_{D^zD^z}(r)$ on the honeycomb lattice as a function of distance $r$ for various iPEPS bond dimensions $D$. (a) Pure Kitaev model ($K=-1, J=0$). (b) Kitaev-Heisenberg model ($K=-1, J=-0.10$). The dashed line indicates the characteristic $1/r^4$ power-law decay of the Kitaev QSL phase, which is clearly observed in both cases.}
\end{figure}

Using the same optimization parameters as for the pure Kitaev model (see TABLE.~\ref{tab:kitaev_data}), we compute the connected dimer-dimer correlation functions for both the pure Kitaev model ($K=-1$) and the Kitaev-Heisenberg model with weak ferromagnetic Heisenberg coupling ($K=-1,J=-0.1$). As shown in Fig.~\ref{fig:dimerdimer}, our results reproduce the characteristic $1/r^4$ decay over a range of distances which systematically extends as the bond dimension $D$ increases. A sufficiently large $\chi$ up to $300$ was used for the evaluation, such that the finite-$\chi$ effects are small compared to the symbol sizes.

\section{Conclusion and Outlook}

We have introduced a QR-based corner transfer matrix renormalization scheme specifically tailored for iPEPS simulations on the honeycomb lattice, with the simulations preserving $\C_{3v}$ symmetry of the wave function. By replacing the eigenvalue decomposition (EVD) or singular value decomposition (SVD) of enlarged corner matrices with local QR decompositions applied to smaller network segments, our implementation, combined with symmetry-projected tensors and virtual bond slicing, achieves substantial improvements in computational speed and memory efficiency while maintaining full compatibility with automatic differentiation. Across representative benchmarks, our method achieves variational accuracies comparable to EVD-based approaches, with more than an order-of-magnitude reduction in wall-clock optimization time and in per-forward iteration cost, without compromising stability or convergence.

These algorithmic advances enable unprecedented precision in iPEPS calculations on the honeycomb lattice, as demonstrated for both the spin-1/2 antiferromagnetic Heisenberg and Kitaev models. For the isotropic Kitaev model, our iPEPS energies extrapolate systematically approach the exact value from $D=4$ to $8$, and the transfer-matrix spectrum analysis reveals a transition from gapped ($D=5$) to gapless ($D\geq6$) behavior, consistent with the known gapless spin liquid nature. Most notably, in the Kitaev-Heisenberg model we find clear numerical evidence for the universal $1/r^4$ algebraic decay of the dimer-dimer correlation function within the spin-liquid regime, demonstrating the persistence of this hallmark signature in the presence of a weak Heisenberg coupling.


While this work focused on ground states, we emphasize that our approach is equally applicable to 2D classical systems at critical point and to finite-temperature quantum systems via the variational tensor network renormalization (VTNR) method~\cite{PhysRevB.98.045110}. The latter is particularly demanding due to the large bond dimensions involved and can therefore greatly benefit from QR-accelerated computations.

Looking beyond the present scope, the general principles of our approach are not restricted to the honeycomb lattice. A compelling future direction involves its extension to other lattices (e.g., triangular, kagome) governed by different point group symmetries. We note, however, that such an extension is non-trivial, as it requires constructing a new set of local mappings—analogous to those defined in Fig.~\ref{fig:1}.

\begin{acknowledgments}
	We thank Illya Lukin, Yuchi He and Honghao Tu for helpful discussions. This project has received funding from the European Research Council (ERC) under the European Union's Horizon 2020 research and innovation programme (grant agreement No.101001604). This work was carried out on the Dutch national e-infrastructure with the support of SURF Cooperative. We acknowledge EuroHPC Joint Undertaking for awarding us access to MareNostrum5 at BSC, Spain under Grant No. EHPC-DEV-2025D06-052.
\end{acknowledgments}

\bibliography{references}

\appendix

\section{$\C_{3v}$ symmetry of the iPEPS for spin-$S$ model on the honeycomb lattice\label{app:c3v}}

The honeycomb lattice inherently possesses $\C_{3v}$ point group symmetry, a fundamental property of its geometric structure. Many physically relevant quantum many-body models defined on this lattice, including the Kitaev-Heisenberg model (the focus of our study), exhibit this geometric symmetry in their Hamiltonians. When a system's Hamiltonian exhibits such symmetries, its ground state is expected to transform according to a direct sum of specific irreducible representations (irrep) of the symmetry group, or to be a superposition of states from such representations if spontaneous symmetry breaking occurs. In the context of iPEPS, imposing this intrinsic geometric symmetry directly on the local tensors is highly beneficial for both accurately representing the physical ground state and achieving computational efficiency during tensor network contraction and optimization.

For a quantum many-body state $\ket{\Psi}$ to possess $\C_{3v}$ symmetry, it must be invariant (up to a global phase) under the action of all symmetry operators $\hat{G}$ belonging to the $\C_{3v}$ point group, i.e., $\hat{G}\ket{\Psi} = e^{i\theta}\ket{\Psi}$. In the iPEPS formalism, this invariance translates into specific constraints on the local tensor $A$ that defines the state. To implement these constraints effectively, it is essential to explicitly define the unitary representations of the symmetry operators acting on the physical degrees of freedom and to understand how they permute the virtual bonds. The choice of the group representation according to which the state transforms dictates the specific form of the unitary representations $U_G$ and the resulting constraints on the tensor. This allows us to construct a variational ansatz that inherently respects the desired symmetry.

The $\C_{3v}$ point group possesses three irreducible representations (irreps): two one-dimensional ($\mathbf{A}_1$ and $\mathbf{A}_2$) and one two-dimensional ($\mathbf{E}$). Any representation of the group, including that acting on the physical Hilbert space, can be decomposed into these irreps. The specific choice of representation (irreducible or a direct sum of irreps) for the iPEPS state dictates the explicit form of the unitary matrices $U_G$ acting on the physical degrees of freedom.

For spin systems on the honeycomb lattice, the $\C_{3v}$ group elements act on the physical legs of the iPEPS tensor through local unitary transformations. For a general spin-$S$ system, where the local Hilbert space dimension is $d=2S + 1$, the spin operators $\mathbf{S} = (S^x, S^y, S^z)$ are represented by $d \times d$ matrices.

The trivial one-dimensional irreps $\mathbf{A}_1$ and $\mathbf{A}_2$ are applicable to certain models, such as the Heisenberg model (after transformed $\pi$-rotation along $y$-axis on B sublattice), where the Hamiltonian exhibits full $\mathrm{SU}(2)$ spin rotation symmetry. In these cases, the physical leg transformations reduce to $U_G = I$ (identity), and symmetry operations are manifested primarily as permutations of virtual bonds. In contrast, for models with bond-dependent anisotropic interactions, such as the Kitaev model, spin operators are explicitly tied to specific lattice directions (e.g., $S_i^\alpha S_j^\alpha$ on a $\alpha$-bond). This spin-lattice locking creates a strong coupling between spin and real space, necessitating a non-trivial representation of the $\C_{3v}$ group to properly encode the symmetry. For the spin-1/2 Kitaev model, the 2D $\mathbf{E}$ irreducible representation is the only valid representation that captures this complex coupling.

\begin{figure}[t]
	\centering
	\includegraphics[width=0.495\textwidth]{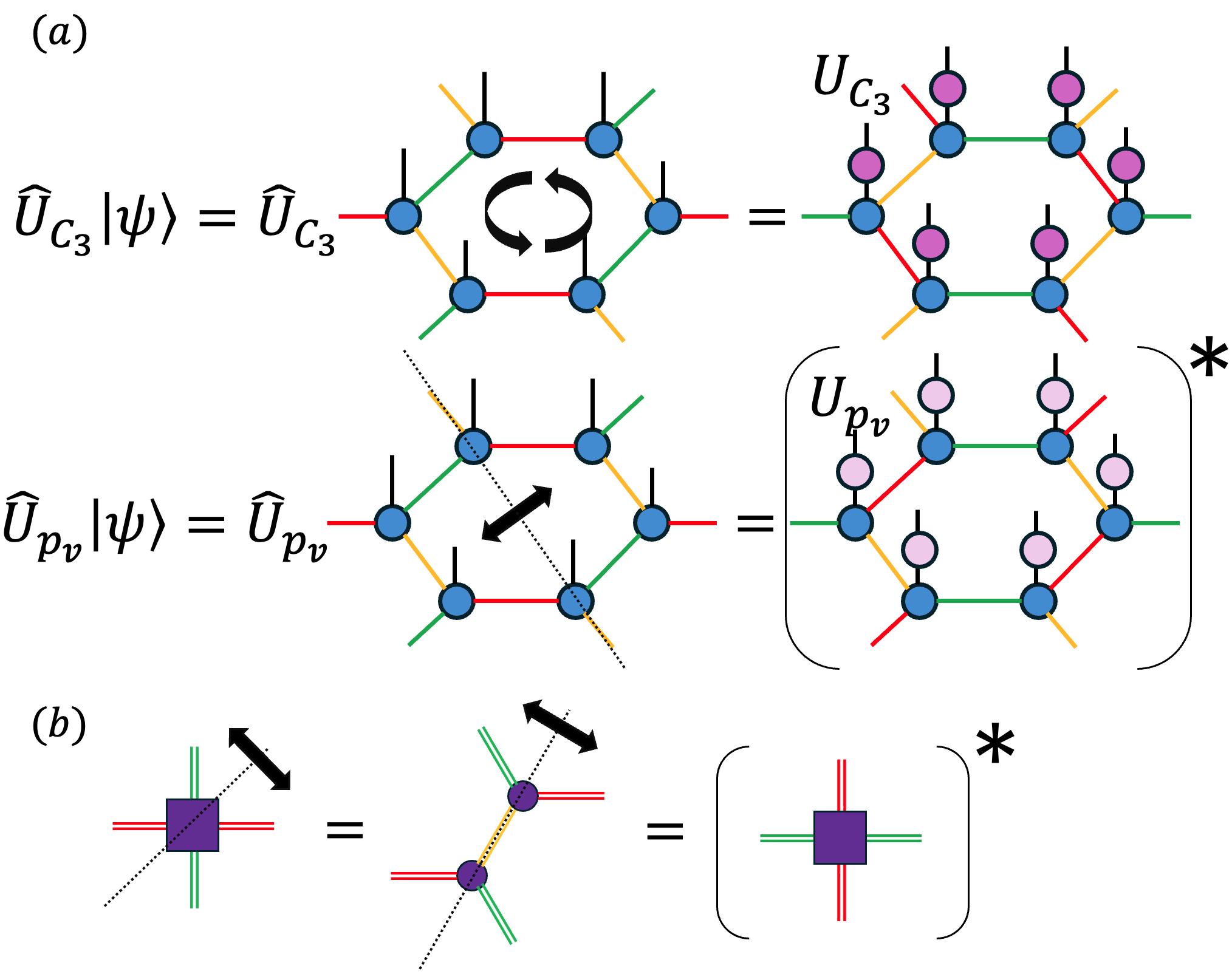}
	\caption{\label{fig:2} Action of $\C_{3v}$ point group operators on the iPEPS for the honeycomb lattice. The asterisk (*) denotes Hermitian conjugation. (a) Action of the three-fold rotation operator $\hat{U}_{\text{C}_3}$ (top) and a mirror reflection operator $\hat{U}_{p,v}$ (bottom) on the iPEPS state $\ket{\Psi}$. Black curved arrows indicate cyclic permutation of virtual bonds, while pink circles represent the application of the group representation matrices ($U_{\text{C}_3}$ or $U_{p,v}$) to the physical legs. (b) Mirror reflection symmetry action on the effective square unit cell tensor obtained through the local mapping procedure.}
\end{figure}

In this $\mathbf{E}$ representation, the symmetry operations act non-trivially on the physical spin degrees of freedom through specific unitary transformations. For the Kitaev model, where the $[111]$ direction serves as the rotational symmetry axis, the three-fold rotation operator $\hat{U}_{\text{C}_3}$ is represented by a $2\pi/3$ rotation around the $\mathbf{n}=(1,1,1)/\sqrt{3}$ axis in spin space:
\begin{equation}
	U_{\text{C}_3} = (-1)^{d-1} e^{i \phi}\exp\left(i \frac{2\pi}{3} \mathbf{n} \cdot \mathbf{S}\right)
	\label{eq:Uc}
\end{equation}
where $e^{i \phi}$ is a flexible phase. For the $\text{E}$ irrep, we typically set $\phi=2\pi/3$ (such that $e^{i 3\phi}=1$). The factor $(-1)^{d-1}$ ensures consistency with the chosen representation for general spin-$S$ systems (for spin-1/2, $d=2$, this factor is $-1$).

For the reflection symmetries, the implementation involves a combination of spatial operations and time-reversal transformations~\cite{Littlejohn2021TimeReversal,PhysRevB.107.054424}. This non-trivial combination arises from spin-lattice coupling, where reflections induce a transformation of spins that includes a time-reversal component, making these reflection operations in spin space anti-unitary.

Specifically, the mirror symmetry operations are constructed using three anti-unitary operators, $\hat{U}_{p_1}$, $\hat{U}_{p_2}$, and $\hat{U}_{p_3}$. These are not independent; for Kitaev model, due to the underlying $\C_3$ rotational symmetry around the $[111]$ axis, defining one allows the others to be derived through group relations. Each operator $\hat{U}_{p_i}$ corresponds to a reflection across a plane whose normal is aligned with a specific direction $\mathbf{m}_i$ in spin space. Their general form on the physical leg is given by:

$$ U_{p_i} = \hat{K} \mathcal{C} U_{p_i}^{\text{rot}} $$

Here, $\mathcal{C}$ denotes the complex conjugation operator, which acts on a state vector by taking the complex conjugate of each of its components. $\hat{K}$ is a $d \times d$ unitary matrix that represents the unitary part of the anti-unitary time-reversal operator. For $d=2$ (spin-1/2 systems), $\hat{K}$ takes the form $i\sigma_y$ (where $\sigma_y$ is the standard Pauli matrix). Its matrix representation in the standard $S^z$ basis is:
$$\hat{K}=\begin{pmatrix}
		0          & 0      & \cdots & 0      & 1      \\
		0          & 0      & \cdots & -1     & 0      \\
		\vdots     & \vdots & \ddots & \vdots & \vdots \\
		(-1)^{d+1} & 0      & \cdots & 0      & 0
	\end{pmatrix}$$

The matrix $U_{p_i}^{\text{rot}}$ is unitary and given by:
\begin{equation}
	U_{p_i}^{\text{rot}} = e^{i \phi_i} \exp\left(i \pi \mathbf{m}_i \cdot \mathbf{S}\right) 
	\label{eq:Up}
\end{equation}
Here, $\exp\left(i \pi \mathbf{m}_i \cdot \mathbf{S}\right)$ represents a $\pi$-rotation in spin space around the axis $\mathbf{m}_i$. The phase factors $e^{i \phi_i}$ should be adapted to ensure consistency with the $\C_3$ rotation relations.

The unit vectors $\mathbf{m}_i$ defining these reflection axes are derived from the $\C_3$ symmetry, which satisfy $\mathbf{m}_i\cdot \mathbf{n}=0$ and are given by:

$$
	\begin{aligned}
		\mathbf{m}_1 & = \frac{1}{\sqrt{2}} \begin{pmatrix} 1, & -1, & 0 \end{pmatrix}, \\[1em]
		\mathbf{m}_2 & = \frac{1}{\sqrt{2}} \begin{pmatrix} 0, & 1, & -1 \end{pmatrix}, \\[1em]
		\mathbf{m}_3 & = \frac{1}{\sqrt{2}} \begin{pmatrix} -1, & 0, & 1 \end{pmatrix}.
	\end{aligned}
$$

These three vectors are oriented at $120^\circ$ to each other. Importantly, all three vectors are orthogonal to the $[111]$ direction. Under a $120^\circ$ rotation around the $[111]$ axis, these three vectors cyclically permute ($\mathbf{m}_1 \to \mathbf{m}_2 \to \mathbf{m}_3 \to \mathbf{m}_1$), reflecting the underlying lattice symmetry. This specific orientation of the reflection axes, along with the inclusion of time-reversal and global phase factors, reflects the underlying $\C_{3v}$ symmetry and its non-trivial influence on the spin degrees of freedom.

These group actions, encompassing both rotation and reflection operations, are visually summarized in Fig.~\ref{fig:2}. As depicted in Fig.~\ref{fig:2}(a), the action of the three-fold rotation operator $\hat{U}_{\text{C}_3}$ on the iPEPS state involves both a cyclic permutation of the virtual bonds connected to each tensor and the application of the local unitary matrix $U_{\text{C}_3}$ to the physical leg. Similarly, the mirror reflection operators $\hat{U}_{p_v}$ involves specific virtual bond exchanges with the application of their corresponding matrices representations $U_{p_v}$ to the physical legs. The black curved arrows in Fig.~\ref{fig:2}(a) indicate the cyclic permutation of virtual bonds, while the pink circles represent the application of the unitary matrices to the physical legs.

With these group operations and their explicit matrix representations defined, we can now construct a $\C_{3v}$-symmetric iPEPS ansatz through symmetry projection. The approach involves summing the local tensor $A$ over all its symmetry-transformed counterparts, effectively projecting it onto the invariant subspace of the $\C_{3v}$ group. This is implemented by applying the following projection operator:
$$ A_{\text{sym}} = \frac{1}{{|G|}}\sum_{g \in \C_{3v}} \mathcal{T}_g (A) $$
where the sum runs over all group elements $g$ in $\C_{3v}$, and $\mathcal{T}_g$ denotes the combined transformation induced by $g$. For unitary operations (rotations), $\mathcal{T}_g(A)$ involves applying the unitary matrix $U_g$ to the physical leg and permuting the virtual indices. For anti-unitary operations (reflections), $\mathcal{T}_g(A)$ involves applying the unitary part $U_g^{\text{rot}}$ to the physical leg, complex conjugating the entire tensor, and permuting the virtual indices.

The enlarged corner tensor obtained from symmetric tensor $A_{\text{sym}}$ possesses Hermicity under reflection as illustrated in Fig.~\ref{fig:2}(b). This rigorous enforcement of symmetry, derived from group representation theory and implemented at the tensor construction level, is a prerequisite for our algorithm. 

\end{document}